\documentclass[journal]{IEEEtran}
\usepackage{graphicx}  
\usepackage{amsmath}
\usepackage{amssymb}
\usepackage{hyperref}
\usepackage{url}
\usepackage{subcaption}

\title{Transferability of data-driven optimization results across multiple pixelated CdZnTe spectrometers}
\author{
    Thomas~D.~MacDonald, Hannah~S.~Parrilla, Jayson~R.~Vavrek
    \thanks{
        T.D.~MacDonald, H.S.~Parrilla, and J.R.~Vavrek are with the Nuclear Science Division, Lawrence Berkeley National Laboratory, Berkeley, CA, 94720 USA.

        This work was performed under the auspices of the U.S.\ Department of Energy by Lawrence Berkeley National Laboratory (LBNL) under Contract DE-AC02-05CH11231.
        The work presented in this paper was funded by the National Nuclear Security Administration of the Department of Energy, Office of International Nuclear Safeguards - Safeguards Technology Development Program.
    }
}
\date{\today}

\begin{document}

\maketitle

\begin{abstract}
Recent work by Vavrek et al.\ (2025) showed that machine learning methods can be used to exploit spatial patterns of performance variations within the highly-segmented H3D M400 gamma spectrometer to improve an overall spectroscopic performance metric.
That work also introduced the {\tt spectre-ml} software, which tests various greedy, heuristic, random, and machine learning clustering algorithms to find the best performing mask for excluding detector regions to improve a user-defined performance metric by training on a given dataset.
In this work, we build off of Vavrek et al.\ (2025) and seek to determine to what extent an optimized binary voxel mask trained on a given dataset can generalize to other datasets.
In particular, this paper evaluates the transferability of masks trained on one M400 dataset to another M400 detector, in order to determine whether the total effort required in designing masks for different detectors and applications can be substantially reduced by using a single common mask. 
It also examines testing and training on different subsets of the same dataset to determine the natural level of variability in optimization results.
In the inter-detector analysis, as expected, the best performing model on each detector is often one trained on that dataset, with an average performance enhancement of $16\%$ when considering the relative uncertainty in a Doniach fit to the $186$~keV peak. 
In comparison, the best transferred masks, with the best on average performance metric across all six detectors, show only a slightly smaller improvement of $13\%$ on average.
These results suggest that high-performing, well-transferable masks can be shared among detectors, reducing or even eliminating the laborious processes of collecting a training dataset and performing the optimization for each detector, ultimately improving safeguards efficiency.

\end{abstract}

\section{Introduction}
The H3D M400 CdZnTe (CZT) detector~\cite{m400_spec_sheet} is a highly segmented, medium energy resolution, room-temperature semiconductor radiation detector that is increasingly used by the International Atomic Energy Agency (IAEA) for non-destructive assay (NDA) in international nuclear safeguards settings~\cite{iaea2022disp, lebrun2022next, dodane2023large}.
While the fine spatial segmentation allows for precise position of interaction localization, irregularities within the CZT crystals and changing proximity to the readout electrodes create large variations in performance (energy resolution and efficiency) among voxels~\cite{bolotnikov2007cumulative, li1999spatial, goodman2022energy}.
Poorly performing voxels can be excluded in order to try to improve spectroscopic performance metrics (e.g., the uncertainty in a peak amplitude, or the resolvability of closely spaced peaks) but at the expense of worse statistics due to fewer counts. 
The optimal selection of voxels would precisely balance these two effects, but cannot be computed directly due to the large number (${\sim}10^{7285}$) of possible voxel combinations, and thus approximate solutions are required.
Previous work~\cite{vavrek2025data} introduced a data-driven approach to finding an approximate solution, {\tt spectre-ml}, which tests various clustering, heuristic, and greedy algorithms to produce masks of active pixels designed to optimize a user-provided metric.
The {\tt spectre-ml} software learns clusters of similarly-performing voxels and tests spectroscopic performance on various voxel cluster combinations, shrinking the space of solutions to search.
This approach is agnostic to the chosen performance metric, which can be computed from single peak fits as in Ref.~\cite{vavrek2025data}, from more complex and application-specific uranium spectrum template fitting code such as {\tt GEM} in Ref.~\cite{vavrek2025bdata}, or more generally from arbitrary end-user analysis code.

As the M400 CZT detector has become more widely deployed, there has been increasing stakeholder interest in the generalizability of a single detector's optimization result to multiple other detector units.
Since the process of collecting training data and running the optimization software can amount to several hours per detector, it would be highly efficient to be able to apply a single optimized mask to many detector units, especially those already in active deployment.
Therefore, in this work, we investigate the extent to which optimization results can be applied across detectors while still retaining high performance.
(We note that some preliminary results were presented in Ref.~\cite{parrilla2025voxel}, but that a number of expansions are included in the present work.)
First, we present inter-detector comparisons of six US National Laboratory M400 detectors in order to characterize the observed level of variation across and within M400 units.
Second, we compute optimized masks from each detector and test their aggregate performance when applied to the entire set of six detectors.
Even despite the observed level of inter-detector variability, we can identify masks that perform well across all detectors, and we quantify the performance loss compared to using a bespoke mask for each detector (i.e., the best mask trained and tested on the same detector).
Finally, we also examine an aspect of intra-detector variability---testing the extent to which subsampling the training or testing data changes the optimized result---which can characterize the overall stability of the optimization.
We find that the optimization results are robust to subsampling, producing similar optimized metric values across most training and testing subsamples.

\section{Methods}\label{sec:methods}

\subsection{Detector optimization overview}\label{sec:opt_overview}
The many-channel-detector optimization technique we seek to characterize the transferability of, {\tt spectre-ml}, was introduced in Ref.~\cite{vavrek2025bdata}, but we provide a short summary here as well.
The method aims to cluster groups of similarly-performing channels (here M400 voxels) so that a relatively smaller number of voxel cluster combinations can be performance tested rather than an infeasible brute-force search over all voxel combinations.
Rather than compute a single fixed performance metric for every voxel (which can be challenging given potentially low statistics at the individual voxel level), we decompose the set of voxel-level spectra using non-negative matrix factorization (NMF)~\cite{lee1999learning, wang2012nonnegative}.
Every voxel spectrum is then approximated by a linear combination of a low (say $n_{\textrm{comp}} = 
1$--$7$) number of non-negative weights and components.
The weights can then be grouped into again a low (say $n_{\textrm{clust}} = 2$--$7$) number of clusters using standard unsupervised algorithms such as agglomerative, Gaussian mixture, or $k$-means clustering.
Alternatively, clusters can be constructed through a class of non-machine-learning greedy algorithms, which compute the performance metric for every detector element at a chosen coarseness (voxel, pixel, depth bin, or crystal) and then accumulate elements by their performance rank.
Finally, clusters can also be defined using various data-independent heuristics, such as grouping depth bins near the anode and pixels near the crystal edges.
The performance metric of interest is then computed on various cluster combinations, and the optimal model is chosen as the one with the best performance metric.
In this way, the method explores many models along the performance tradeoff between including poorly performing voxels and their associated efficiency gain, allowing the chosen performance metric to select the optimal balance.

In this work, the chosen performance metric is the relative uncertainty in the Doniach fit amplitude of the $186$~keV U-235 series peak, where the Doniach peak shape is given by~\cite{doniach1970many, hyperspy_doniach}
\begin{align}\label{eq:doniach}
    f(E; \mu, A, \sigma, \gamma) &= \frac{A \cos\left[ \pi \gamma/2 + (1-\gamma) \tan^{-1}(\epsilon / \sigma) \right]}{\left( \sigma^2 + \epsilon^2 \right)^{(1-\gamma)/2}},\\
    \epsilon &\equiv E - \mu + \sigma \cot\left( \frac{\pi}{2-\gamma}\right).
\end{align}
Here $E$ is the photon energy, $\mu$ is the peak centroid, $A$ is the peak amplitude, $\sigma$ is the Gaussian component of the peak width, and $\gamma$ is the peak tailing or asymmetry component of the width.
A linear background term is also typically incorporated in the fit.
In this work we use a fit energy ROI of $186 \pm 40$~keV, fitting only the main $186$~keV photopeak.
We do not mask out the smaller peaks in the shoulders of this fit, since they appear to have little effect on the background fit accuracy, but this could be done in future studies.
We also use the counts in this energy ROI to estimate the $186$~keV efficiency relative to bulk after voxel selections are applied.
Although the integral of Eq.~\ref{eq:doniach} diverges~\cite{casaxps_manual, evans1991curve}, the Doniach model accurately fits CZT peak shapes, and thus the relative uncertainty in $A$ can be a useful proxy for improving downstream metrics such as enrichment precision.
We emphasize again that more specific performance metrics can be chosen by downstream users (see, e.g., Ref.~\cite{vavrek2025bdata}).

\subsection{Datasets}
The transferability of the method's results across different detectors is studied here primarily using the multi-detector uranium standard datasets from the Gamma Rodeo project~\cite{smith2024summary}, which aimed to benchmark the M400's performance for various uranium enrichment measurements and software tools.
Six M400 units from different US National Laboratories (Brookhaven, Idaho, Los Alamos, Oak Ridge, Pacific Northwest, and Sandia National Laboratories) were individually and reproducibly placed in a shielded and collimated setup in which the entire detector field of view was covered by a uranium enrichment standard source, and run for an equal dwell time of $30$~minutes.
The consistent source and geometry for each detector allows for direct comparisons of detector performance and subsequent testing of binary voxel mask transferability.
Sources used in this work include the $1.94\%$-enriched Standard Reference Material (SRM) 969~\cite{srm969} and the $20.11\%$-enriched Certified Reference Material (CRM) 146~\cite{crm146}.
We note that in one of the latter source measurements, the INL measurement ended after only ${\sim}1000$~s, so inter-detector comparisons are time-normalized where necessary.
Unless otherwise stated, inter-detector variation characterizations are performed using gross counts rather than any specific photopeak energy region of interest (ROI), but because one of the six detectors (SNL) has a higher low energy threshold than the other five, we consider the gross counts ${\geq}50$~keV (see the later Fig.~\ref{fig:spectra_pairs_194}).

In the inter-detector characterization study, we also use an additional dataset consisting of three measurements of an Eu-154 check source performed with LBNL M400 detectors---two denoted ``Fleming'' and ``Kornhauser'', and one ``Loaner'' unit provided by the vendor while Fleming was under repair.
These measurements were part of the initial performance checks performed upon receipt of each detector, rather than simultaneous measurements of the source, but still provide a valuable expansion in the number of available inter-detector comparison datasets.

\subsection{Analysis methods}\label{sec:methods_analysis}
To characterize spatial performance variations both within and across M400 units, we reduce the 3D listmode count data in two different ways, namely a 2D heatmap of gross count rate in pixel space (summed over depth bin) and a 1D plot of gross count rate in depth bin (summed over pixels).
Detectors are compared by computing count rate ratios among all detector pairs, and deviations from unity are noted.

In the inter-detector transferability studies, for each detector dataset, the listmode data were randomly divided into train and test sets with an 80:20 split.
We applied the {\tt spectre-ml} code as described in Section~\ref{sec:opt_overview}, evaluating greedy, heuristic, and clustering algorithms with the hyperparameters and other configuration settings shown in Table~\ref{tab:hyperparam}.

\begin{table}[!htbp]
\centering
\caption{Inter-detector transferability parameters}
\label{tab:hyperparam}
\begin{tabular}{c|c}
\textbf{Parameter} & \textbf{Values} \\
\hline
energy {[}keV{]} & 186 $\pm$ 40 \\
$n_{\textrm{clust}}$ & 3,4,5,6 \\
$n_{\textrm{comp}}$ & 3,4,5,6 \\
test fraction & 0.2 \\\hline
ML clusterers & agglomerative, Gaussian mixture \\
greedy clusterers & voxel, depth bin \\
random clusterers & voxel, depth bin, pixel \\
heuristic clusterers & equal depth, edge and anode \\
\end{tabular}
\end{table}

For each training dataset, a set of many not-necessarily unique masks was created by the model selection process.
The same parameter space was searched for each detector, but differences in the training data can result in different masks.
Each set of masks was evaluated against every test set using the Doniach amplitude relative uncertainty metric for the $186$~keV peak, giving six sets of models, each tested against six test sets.
For each set of self-tested models, e.g., train and test sets are drawn from the same detector, the best performing mask was identified and the metric value noted.
The metric of the bulk (i.e., all voxels on) for each test set was recorded.
The best performing transferred masks were found by calculating each mask's mean metric across all test sets.

The intra-detector tests involve subsampling either the training or testing dataset to observe the stability of optimization outputs.
In the testing set subsample analysis, each count in the test listmode data was randomly assigned to one of five subsamples.
The training set was held constant and used to compute an optimal mask from the entire training set for each detector.
The top $10$ trained models for each of the six detectors were evaluated against all five of the subsamples, including those from different detectors, resulting in $1800$ different evaluation combinations across all test/train pairs.
The mean and standard deviation of the metrics of the subsamples of each lab's dataset were calculated and plotted.

Similarly, to evaluate the variability in the transferability of masks trained on different and smaller training sets, the training data were randomly divided into five training subsamples.
The full test sets were used to evaluate all of the training subsamples.
Since the training data were different, the same masks would not necessarily be present in each subsample for random and data-dependent mask types (e.g. Agglomerative Clustering, Gaussian Mixture, Random Depth Bin, etc.).
Since masks of these types could not be directly compared between the subsamples, instead the best performing (lowest metric) mask of each type was compared and the means and standard deviations of those masks calculated.
Heuristic masks (e.g. edge and anode, equal depth, and bulk) are found in a non-random and non-data dependent way, were found by all subsamples and since the test set was constant, are only presented once.

\section{Results}\label{sec:results}

\subsection{Inter-detector variation characterization}\label{sec:results_characterization}
\begin{figure*}[!htbp]
    \centering
    \includegraphics[width=1.00\linewidth]{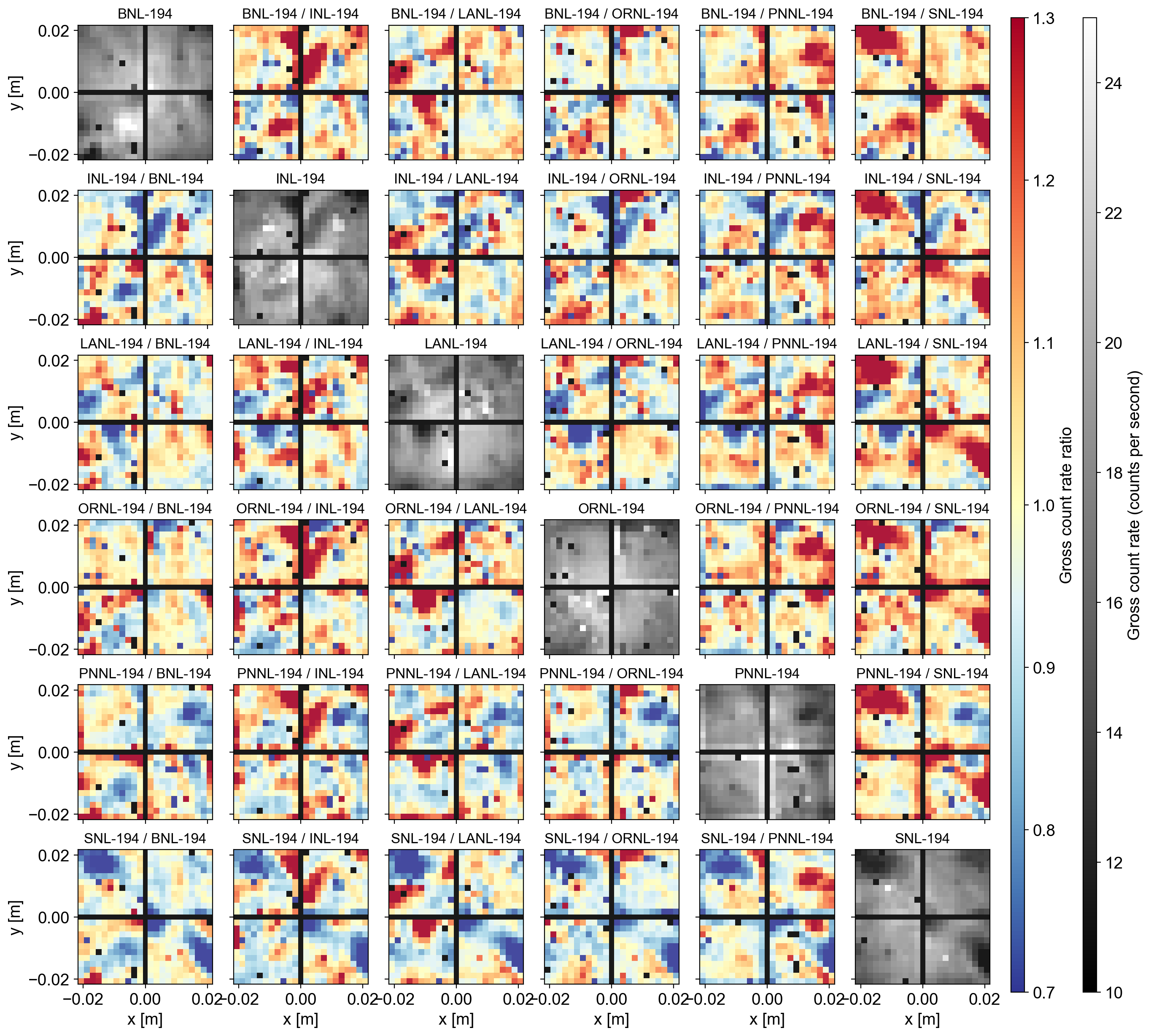}
    \caption{
        Heatmaps of gross count rate data from the $1.94\%$-enriched uranium standard in each of the six M400 detectors.
        Diagonals (grayscale): absolute gross count rates, in counts/s.
        Off-diagonals (blue-yellow-red): ratios of gross count rates between each detector pair, aggregated by detector pixel.
        The central black $+$ is the gap between the four CZT crystals while the individual black pixels are dead pixels with no recorded counts.
    }
    \label{fig:heatmap}
\end{figure*}

\begin{figure}[!htbp]
    \centering
    \includegraphics[width=1.0\linewidth]{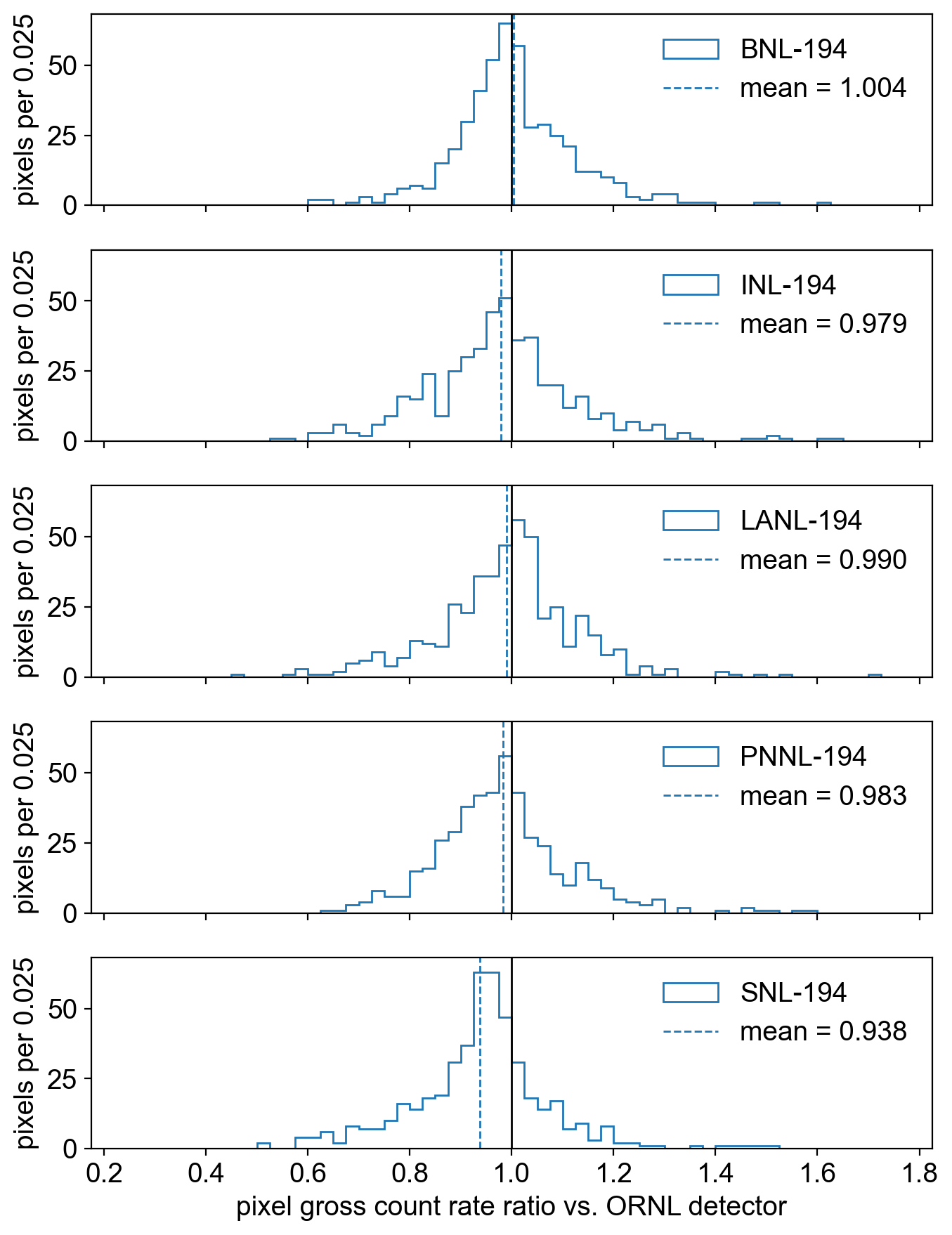}
    \caption{Distribution of per-pixel gross count rate ratios in Fig.~\ref{fig:heatmap}, normalized to the ORNL dataset.}
    \label{fig:ratio_distribution}
\end{figure}

\begin{figure}
    \centering
    \includegraphics[width=1.0\linewidth]{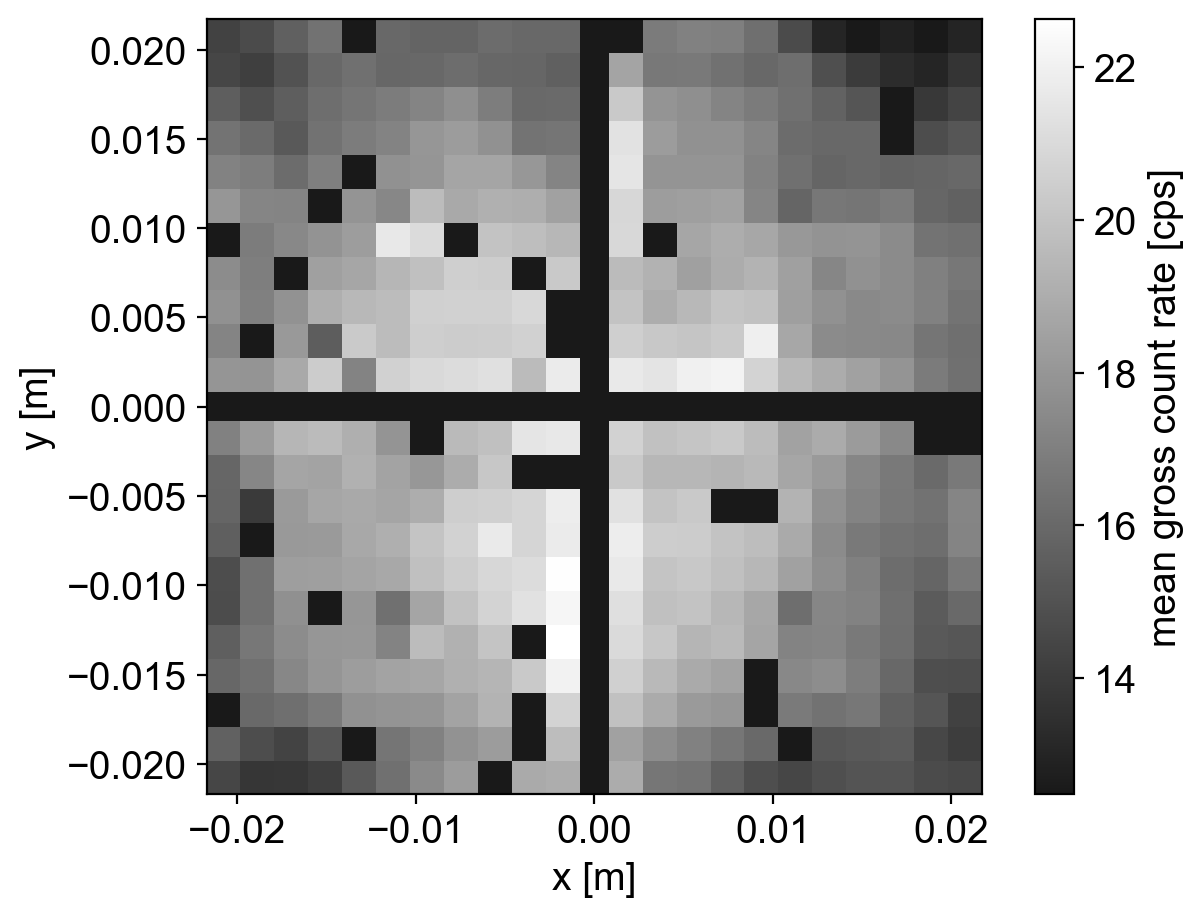}
    \caption{
        Gross count rate per pixel, averaged over all six detectors.
        Pixels that are dead in any detector are set to black.
        As in Fig.~\ref{fig:heatmap}, the central black $+$ is the gap between the four crystals.
    }
    \label{fig:heatmap_gross_count_rate}
\end{figure}

Fig.~\ref{fig:heatmap} shows the pixel-wise variation in gross count rate among the six M400 detectors tested, using the $1.94\%$-enriched uranium standard.
In general, there are substantial (${\lesssim}30\%$) variations in gross count rate per pixel both within and across detectors---see Fig.~\ref{fig:ratio_distribution}.
The spatial extent of the variations is typically several pixels across, or as large as ${\sim}1$~cm.
Fig.~\ref{fig:heatmap_gross_count_rate} shows the gross count rate pixel heatmap averaged over all six detectors.
The spatial variations observed in Fig.~\ref{fig:heatmap} have largely averaged out, though there is still a reduction in rates towards the outer corners of the detectors.

\begin{figure}[!htbp]
    \centering
    \includegraphics[width=1.0\linewidth]{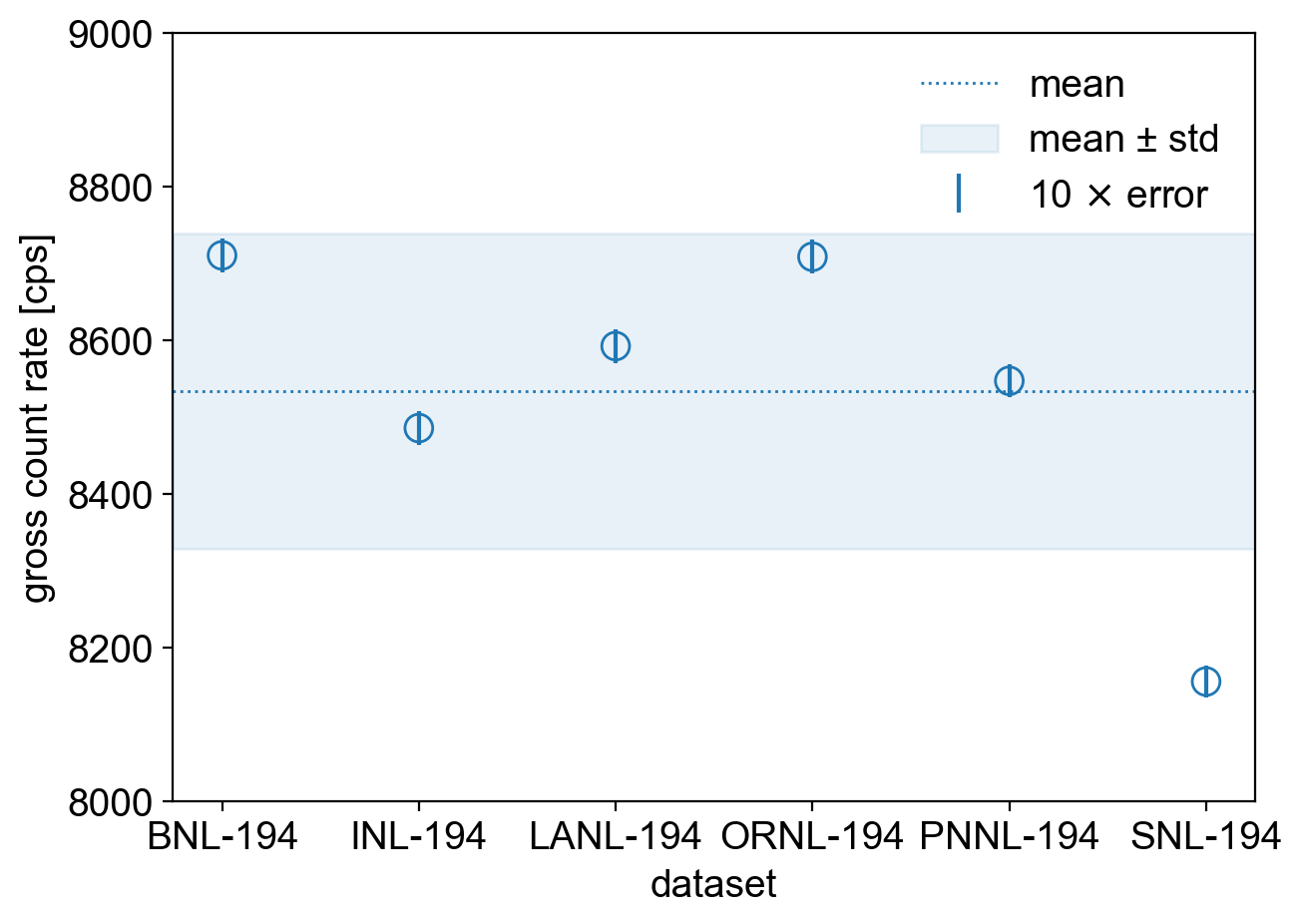}
    \caption{
        Absolute gross count rates in each detector dataset.
        The error bars ($1\sigma$) are multiplied by $10 \times$ for visibility.
        The horizontal band shows the sample standard deviation of the six rates about the sample mean.
    }
    \label{fig:gross_count_rates}
\end{figure}

\begin{figure}[!htbp]
    \centering
    \includegraphics[width=1.0\linewidth]{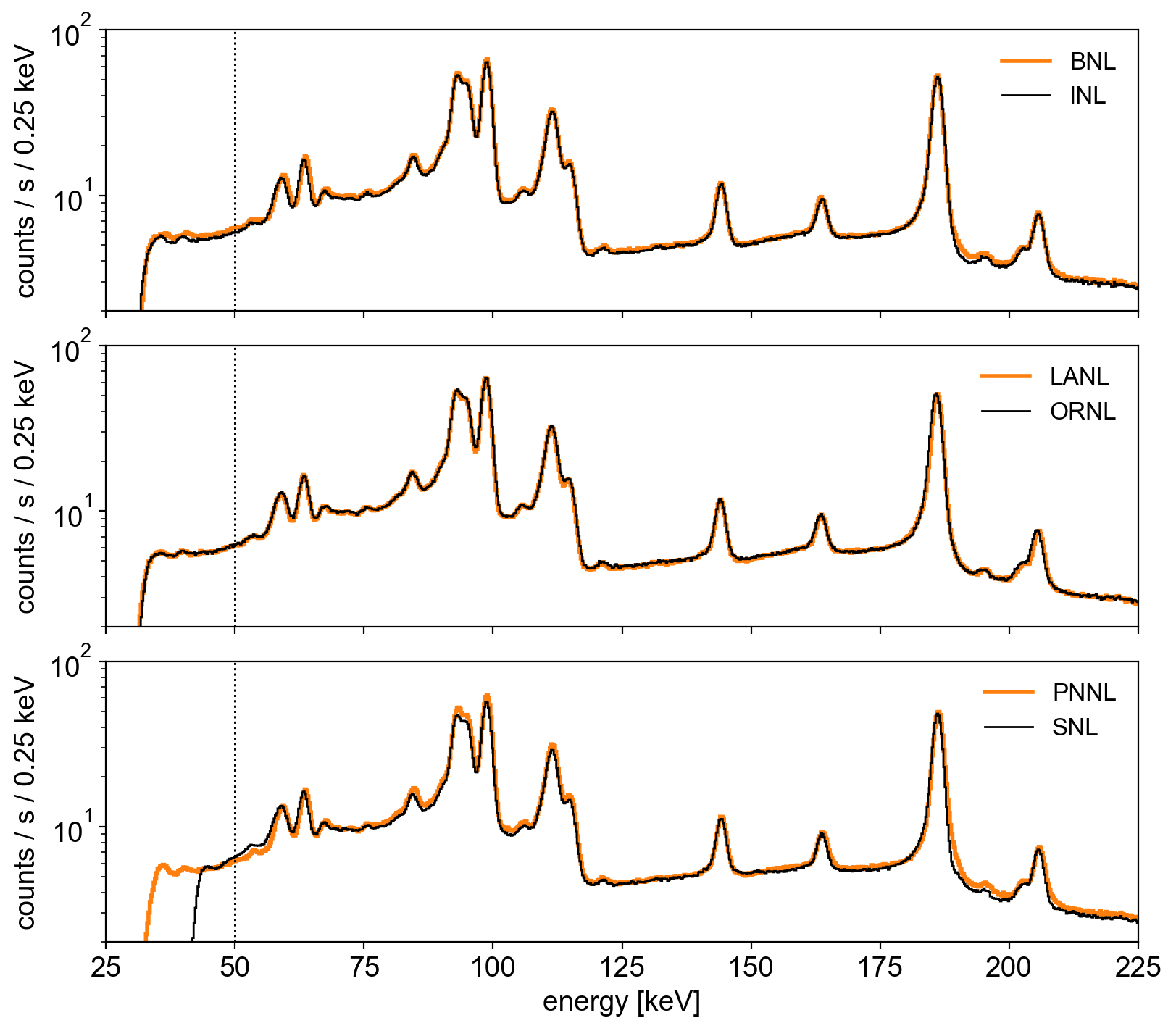}
    \caption{
        Realtime-normalized spectra from the $1.94\%$-enriched uranium standard in each of the six M400 detectors, zoomed to the low-energy region including the $186$~keV U-235 peak.
        The dotted line at $50$~keV indicates the lower energy threshold used throughout this study.
        Spectra are grouped into three pairs solely for visual clarity.
    }
    \label{fig:spectra_pairs_194}
\end{figure}

Fig.~\ref{fig:gross_count_rates} shows the total absolute gross count rates in each detector.
Despite the ${\lesssim}30\%$ \textit{spatial} variations in count rate observed in Fig.~\ref{fig:heatmap}, the \textit{total} count rates summed across all detector elements are consistent to within a sample standard deviation of $2.4\%$.
The SNL detector is the largest outlier with a $4.4\%$ deviation in gross count rate from the mean, even after applying the $50$~keV threshold correction.
A similar deviation vs.\ the ORNL detector can also be seen in Fig.~\ref{fig:ratio_distribution}.
Fig.~\ref{fig:spectra_pairs_194} shows the energy spectra in the six detectors, where again the datasets are largely consistent with each other.

\begin{figure}[!htbp]
    \centering
    \includegraphics[width=1.0\linewidth]{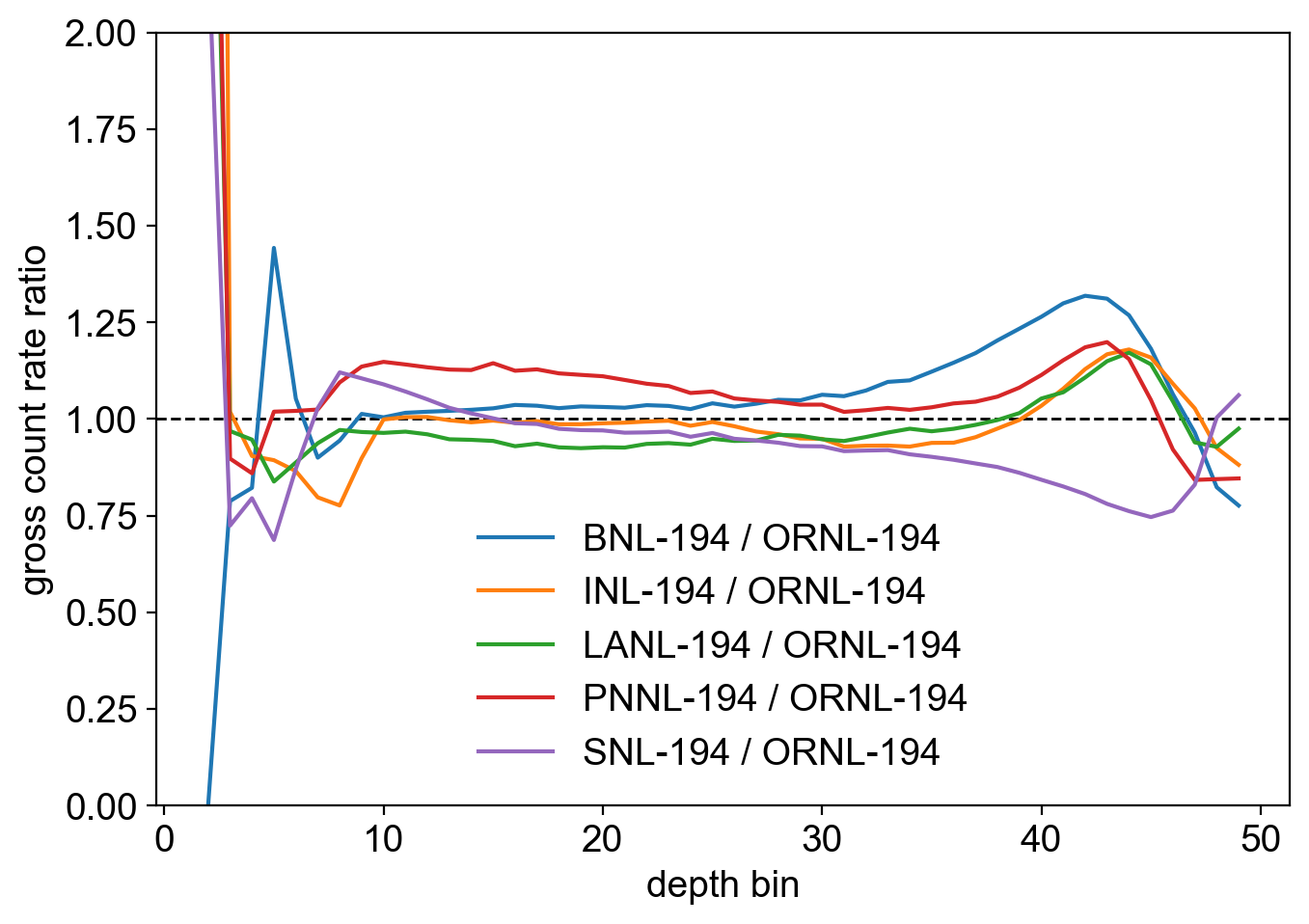}
    \caption{
        Gross count rate vs.\ depth bin from the $1.94\%$-enriched uranium standard in five of the M400 detectors, normalized to the ORNL detector.
        Depth bin $0$ is nearest the anode at the back of the detector while depth bin $50$ is nearest the cathode at the front exposed face of the detector.
    }
    \label{fig:count_rates_vs_depth_bin}
\end{figure}

Fig.~\ref{fig:count_rates_vs_depth_bin} shows a similar count rate comparison across detectors, except performed in the detector depth dimension.
For brevity, only curves normalized against the ORNL detector are shown.
Here the variations are smaller than in pixel space, especially within the central detector volume.

\begin{figure}[!htbp]
    \centering
    \includegraphics[width=1.0\linewidth]{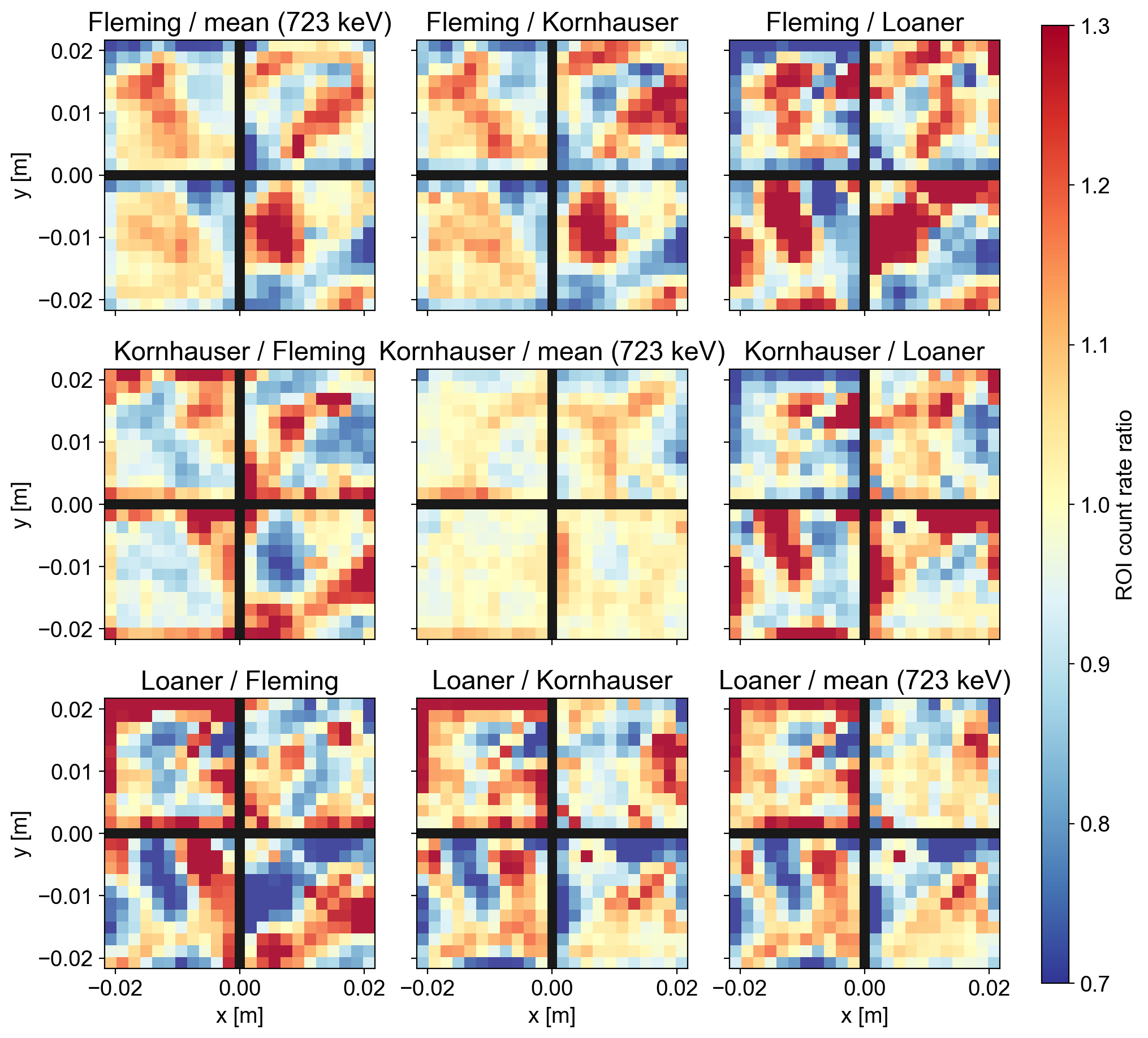}
    \caption{
        Heatmaps of relative count rate data from the Eu-154 $723$~keV ROI in the three LBNL M400 detectors.
        Diagonals: ratio of counts in a pixel to the mean per-pixel counts in that detector.
        Off-diagonals: pixel-wise ratio between detectors.
    }
    \label{fig:relative_heatmaps_723}
\end{figure}

\begin{figure}[!htbp]
    \centering
    \includegraphics[width=1.0\linewidth]{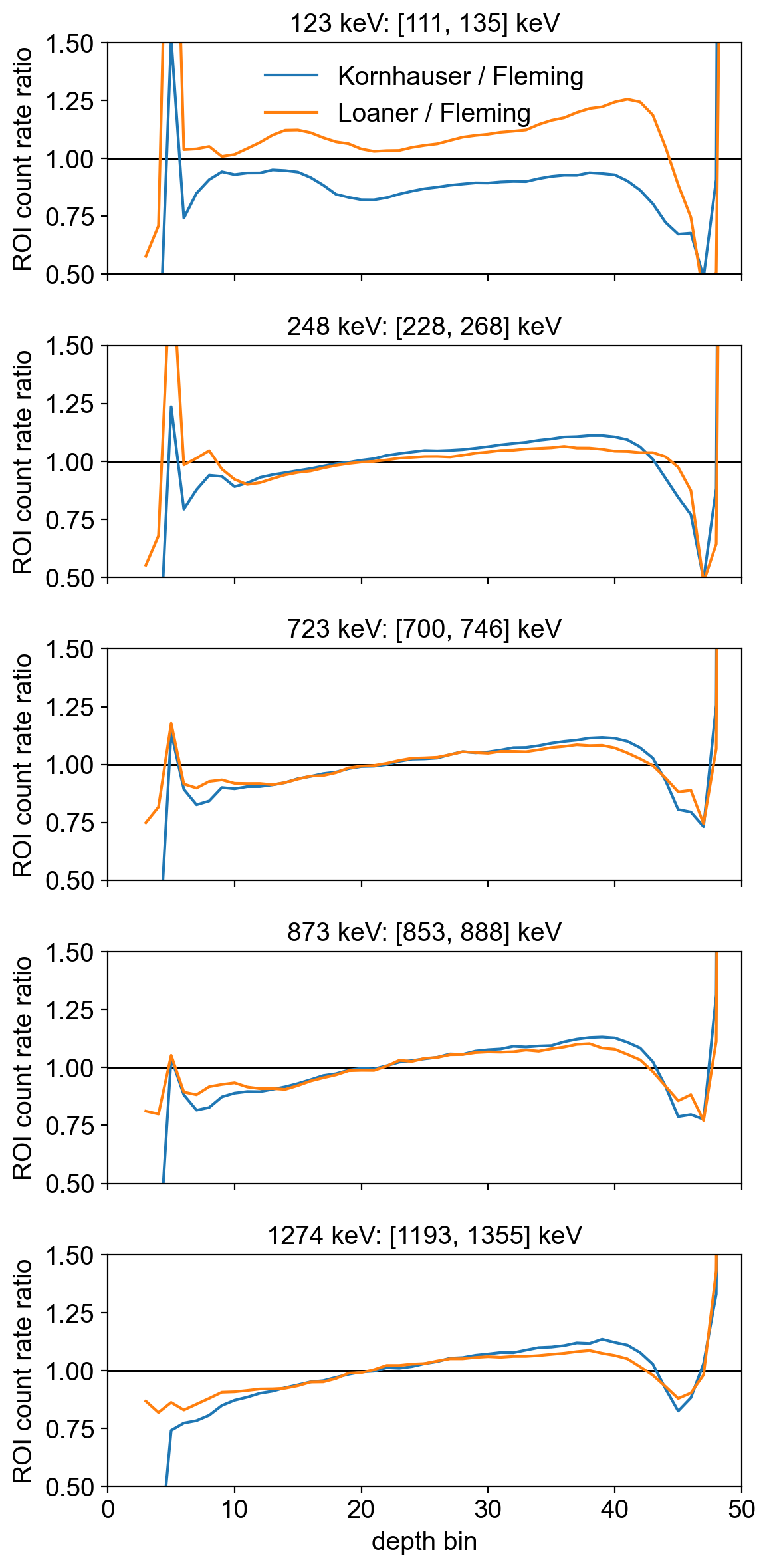}
    \caption{
        Eu-154 ROI count rate ratios in the three LBNL M400 detectors.
    }
    \label{fig:depth_ratios}
\end{figure}

Finally, Figs.~\ref{fig:relative_heatmaps_723} and \ref{fig:depth_ratios} show similar analyses for the Eu-154 measurements with the three LBNL M400 detectors.
Here, analyses are performed on Eu-154 photopeak ROIs rather than gross counts.
In particular, Fig.~\ref{fig:relative_heatmaps_723} shows the pixel-wise heatmap of count rate ratios in the $123$~keV peak.
Since the geometry was less well-controlled than in the uranium measurements, the ROI counts in each pixel are normalized to the mean pixel ROI counts before comparing across detectors.
Similar to the six-detector uranium gross count rate dataset in Fig.~\ref{fig:heatmap}, ${\lesssim}30\%$ variations are present at the scale of ${\sim}1$~cm both within and across detectors, although the variations within the Kornhauser unit are substantially less pronounced.
Fig.~\ref{fig:depth_ratios} shows a similar analysis in the depth dimension of the detectors, normalized against the Fleming unit.
The ratio variations over depth bin are consistent with the gross counts analysis Fig.~\ref{fig:count_rates_vs_depth_bin}, with stronger differences between Kornhauser and the Loaner unit observed at the lowest energy of $123$~keV.

\subsection{Inter-detector transferability}\label{sec:results_transferability}

Given the characterization of intra- and inter-detector differences in Section~\ref{sec:results_characterization}, we now characterize the extent to which voxel selection models can be transferred across detectors despite this variability.

\begin{figure*}[!htbp]
    \centering
    \includegraphics[width=1.0\linewidth]{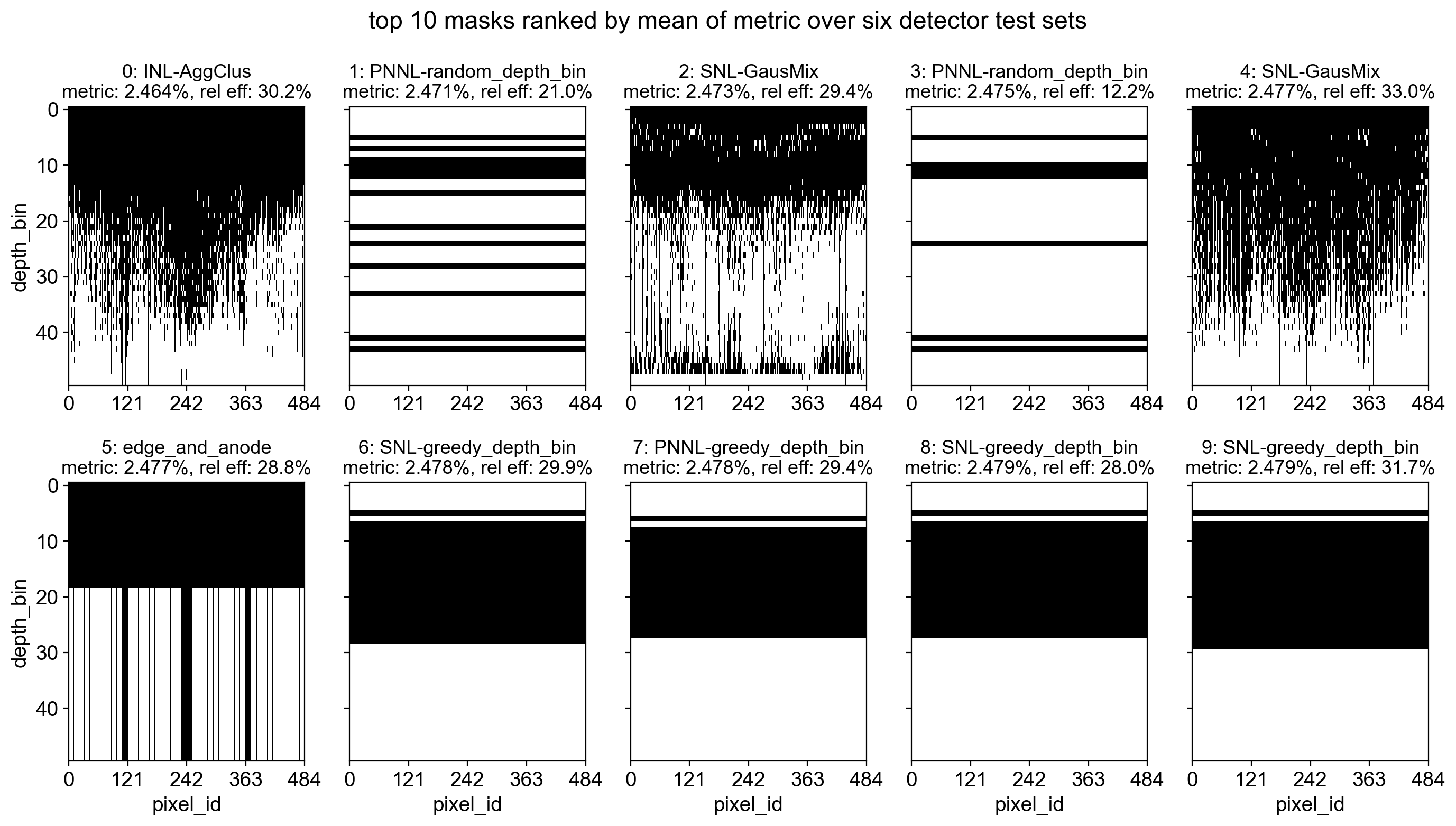}
    \caption{
        Top 10 transferred masks based on mean metric calculated over all test sets.
        Black indicates that the voxel is included in the mask, white indicates that it is excluded. 
        The clustering algorithm that generated the mask and the mean metric over all test sets is shown above each mask, along with an estimate of the efficiency relative to the bulk (unoptimized) detector in the training set energy ROI.
        For masks that depend on the training data (all but the heuristic class), the training set detector is specified in the title.
        The three SNL greedy depth bin masks differ by one or two depth bins upon close inspection.
    }
    \label{fig:masks_top}
\end{figure*}

\begin{figure}[!htbp]
    \centering
    \includegraphics[width=1.0\linewidth]{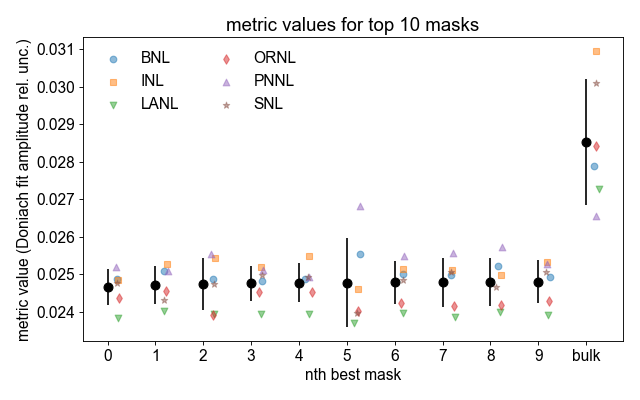}
    \caption{
        Metric comparison between the top 10 transferred masks and bulk.
        Colored points show individual metrics calculated for each detector test set.
        Black points (bars) show the mean (standard deviation) metric over all six detector test sets, and correspond to each subplot in Fig.~\ref{fig:masks_top}.
        The individual detector points are given a small offset and jitter for visual clarity.
    }
    \label{fig:top_metrics}
\end{figure}

\begin{figure}[!htbp]
    \centering
    \includegraphics[width=1.0\linewidth]{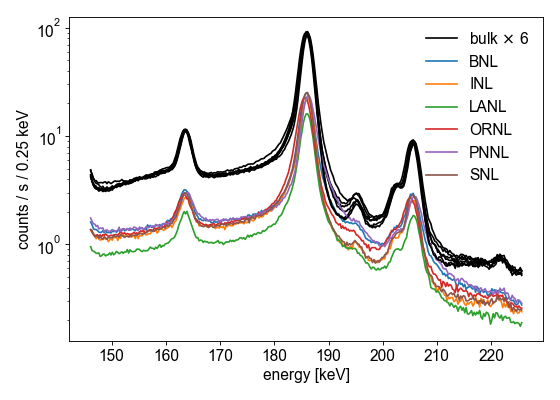}
    \caption{
        Test set spectra from the $20.11\%$-enriched uranium standard after applying the top (rank 0) mask of Fig.~\ref{fig:masks_top}.
    }
    \label{fig:rank0_topspectra}
\end{figure}

\begin{figure}[!htbp]
    \centering
    \includegraphics[width=1.0\linewidth]{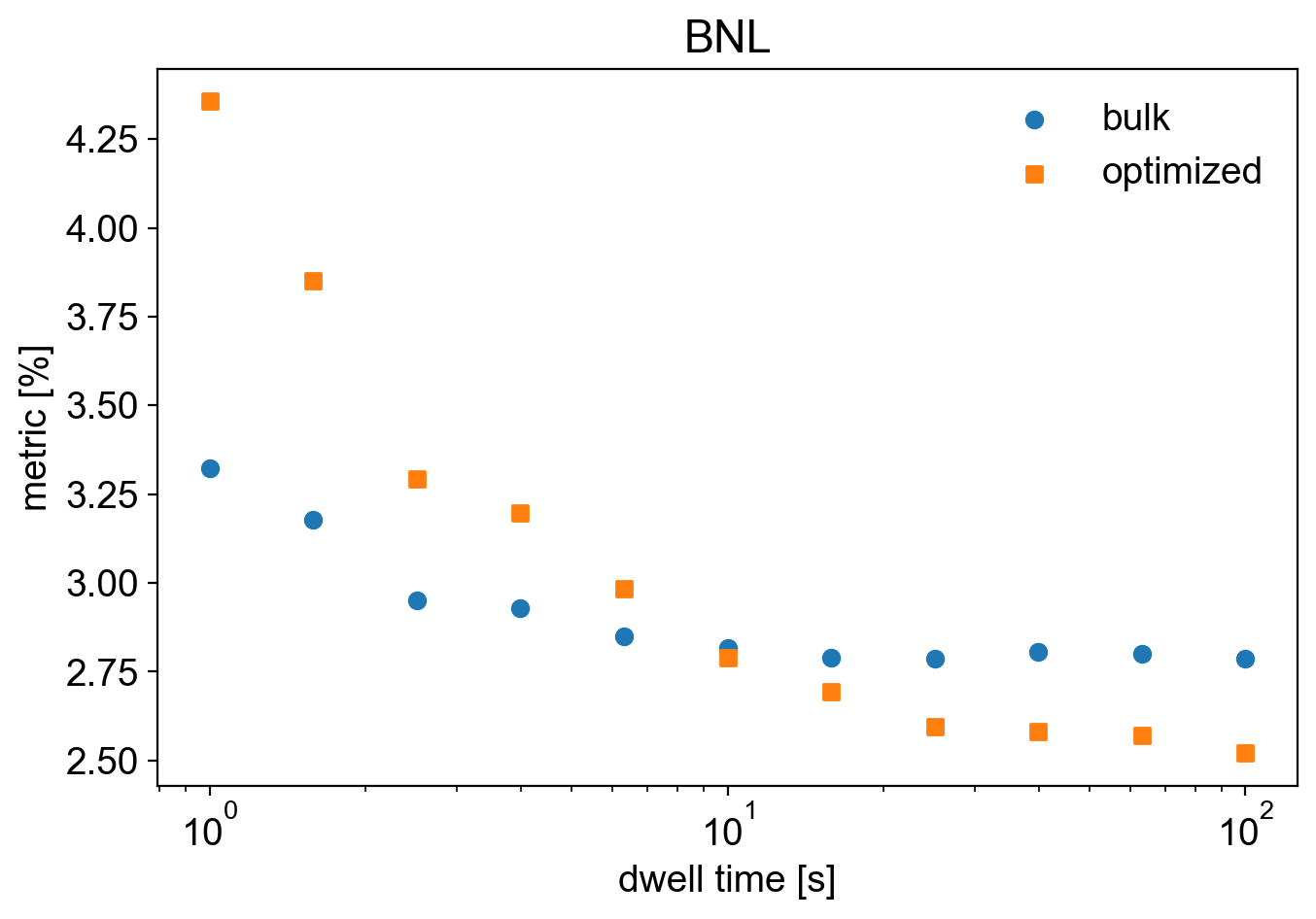}
    \caption{
        Doniach amplitude uncertainty metric vs.\ dwell time for the BNL detector using the bulk (blue circles) and best transferred mask (orange squares).
    }
    \label{fig:metric_vs_dwell_time}
\end{figure}

As described in Section~\ref{sec:methods_analysis}, candidate masks are found by training the various algorithm classes (ML, greedy, heuristic) on the six detectors using the $20.11\%$-enriched measurements, and then evaluated by their mean Doniach fit amplitude relative uncertainty metric over the six testing sets.
Fig.~\ref{fig:masks_top} shows the top 10 masks, as ranked by their mean metric over the six detectors.
Aside from the random depth bin clusterers, most of the masks show similar selection of low-depth-bin values but may arrive at those masks via different clustering algorithms.
The best-performing mask is generated by training on the INL detector, despite its aforementioned shorter dwell time of $1013$~s instead of $1800$~s, using $5$ NMF components and removing $3$ of $4$ agglomerative clusters.
This mask achieves an average Doniach amplitude relative uncertainty of $2.46\%$ compared to the average bulk metric of $2.85\%$.
Fig.~\ref{fig:top_metrics} shows the distribution of metric values of the masks, both with and without averaging over test set detectors.
Each of the top 10 masks outperforms the bulk model both on average and, with the exception of only one outlier (PNNL, $n=5$), in every detector test set, indicating that choosing one of the top 10 transferred masks can reliably improve detector performance across multiple detectors.
Fig.~\ref{fig:rank0_topspectra} shows the test set spectra after applying the best (rank 0) transferred mask.
Applying this mask reduces five of the six test datasets to $30$--$36\%$ of their relative efficiency compared to their bulk datasets, and the LANL dataset to $23\%$.
Finally, Fig.~\ref{fig:metric_vs_dwell_time} shows the evolution of the Doniach amplitude uncertainty metrics as a function of dwell time on the BNL detector, with and without the best transferred mask.
The bulk (unoptimized) metric quickly saturates as the statistical counting uncertainty decreases and the systematic fit error starts to dominate, while the reduction of the latter component in the optimized model allows the total uncertainty to decrease with further dwell time.

\begin{figure*}[!htbp]
    \centering
    \includegraphics[width=1.0\linewidth]{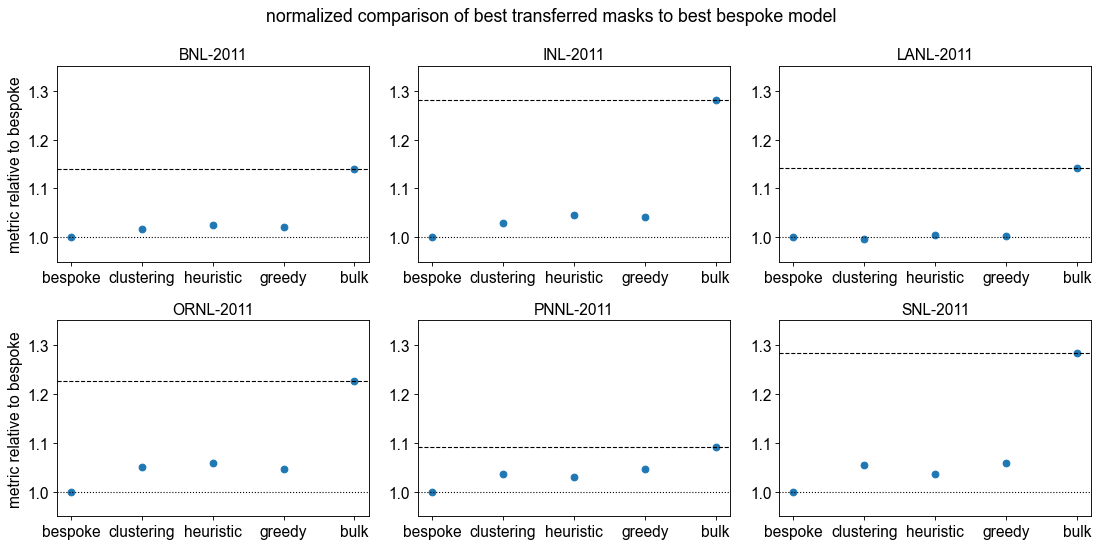}
    \caption{
        Performance (lower values are better) of transferred ML clustering, heuristic, greedy, and bulk masks, relative to the bespoke model for each test set.
        The dashed horizontal line shows the performance of the bulk mask and the dotted horizontal line shows the performance of the bespoke mask for comparison.
        The non-bespoke, non-bulk masks are the top performing masks by mean metric over all test sets for each clustering type.
        The bulk mask does not exclude any voxels.
    }
    \label{fig:bespoke_performance}
\end{figure*}

Fig.~\ref{fig:bespoke_performance} further quantifies the performance loss of the transferred models relative to the bespoke model for each of the six detectors.
In all detectors, the bulk (unoptimized) model is outperformed by the best result from each class of clusterers.
When transferred across detectors, masks typically (with the exception of the LANL detector) approach but do not outperform the bespoke masks.
Relative to the bulk models, the best transferred mask provides an $13\%$ average improvement, while bespoke masks provide $16\%$ on average.
The difference between these two model classes is typically smaller than the difference to the bulk (unoptimized) results, again indicating that transferable models can be found that improve performance relative to bulk across multiple detectors, but that small further performance gains can of course often be obtained via the bespoke masks.

\subsection{Subsampling analysis}\label{sec:results_subsampling}

\begin{figure*}[!htbp]
    \centering
    \includegraphics[width=1.0\linewidth]{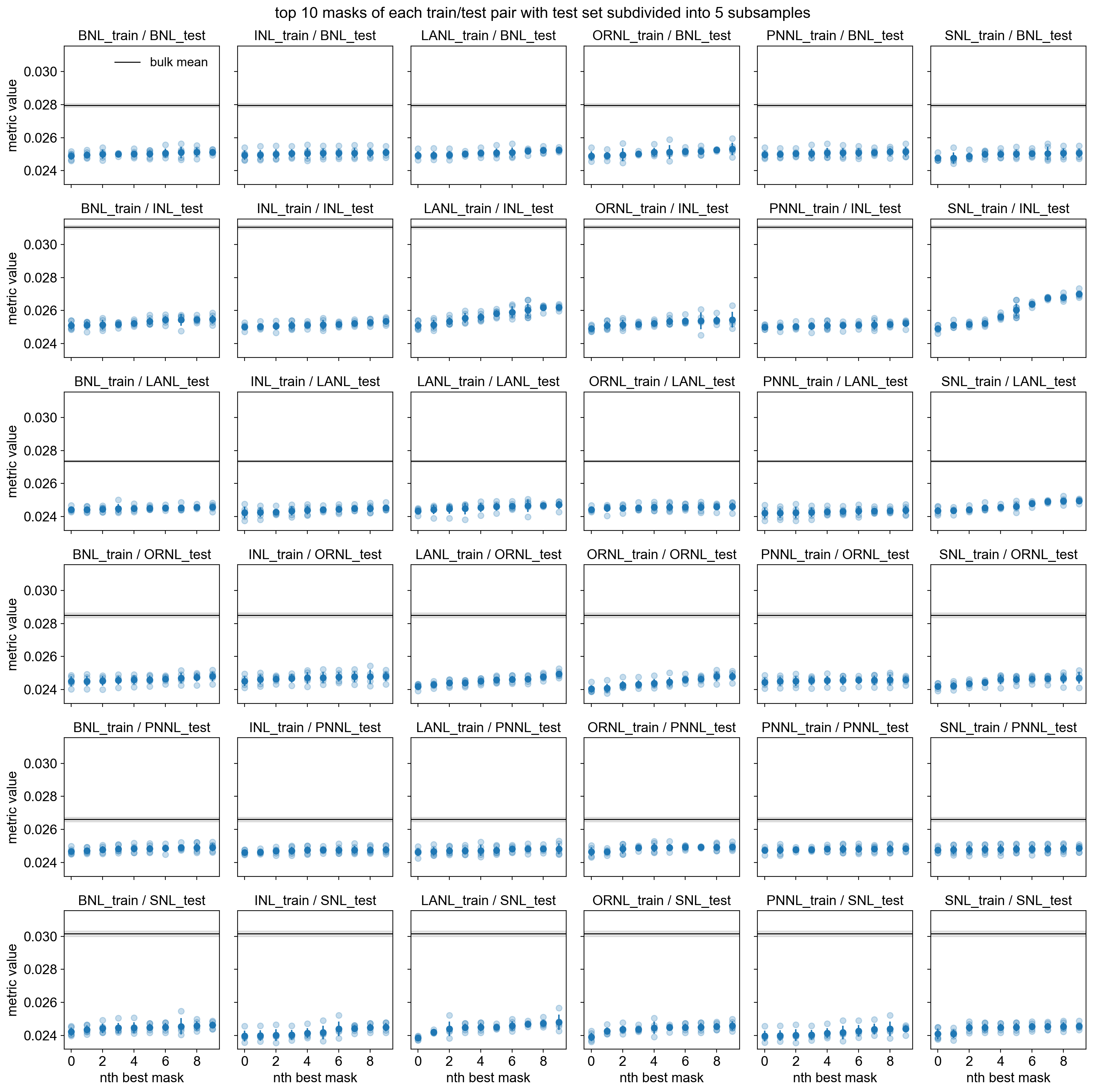}
    \caption{
        Metrics of the top 10 masks from each combination of training/testing dataset, where the testing set has been divided into five subsets.
        Each column indicates a constant training set, each row a constant test set.
        Each translucent point is the metric from a single test subset, while the opaque point indicates the mean of the metric over the five subsets, and bars (where visible) indicate the standard deviation of the five metrics.
        The black line and gray error band indicate the mean and standard deviation of the bulk (unoptimized) metric over the five subsamples of that test set, which is constant in each row.
    }
    \label{fig:subtest}
\end{figure*}

Fig.~\ref{fig:subtest} shows the results of the test set subsampling analysis.
The variation in metric values across the five subsamples is small in nearly all tests, indicating stability against both random variations in the training set and the associated $5\times$ reduction in training dataset size.
Even the worst-performing subsamples reliably perform much better than the bulk (unoptimized) values.

\begin{figure*}[!htbp]
    \centering
    \includegraphics[width=1.0\linewidth]{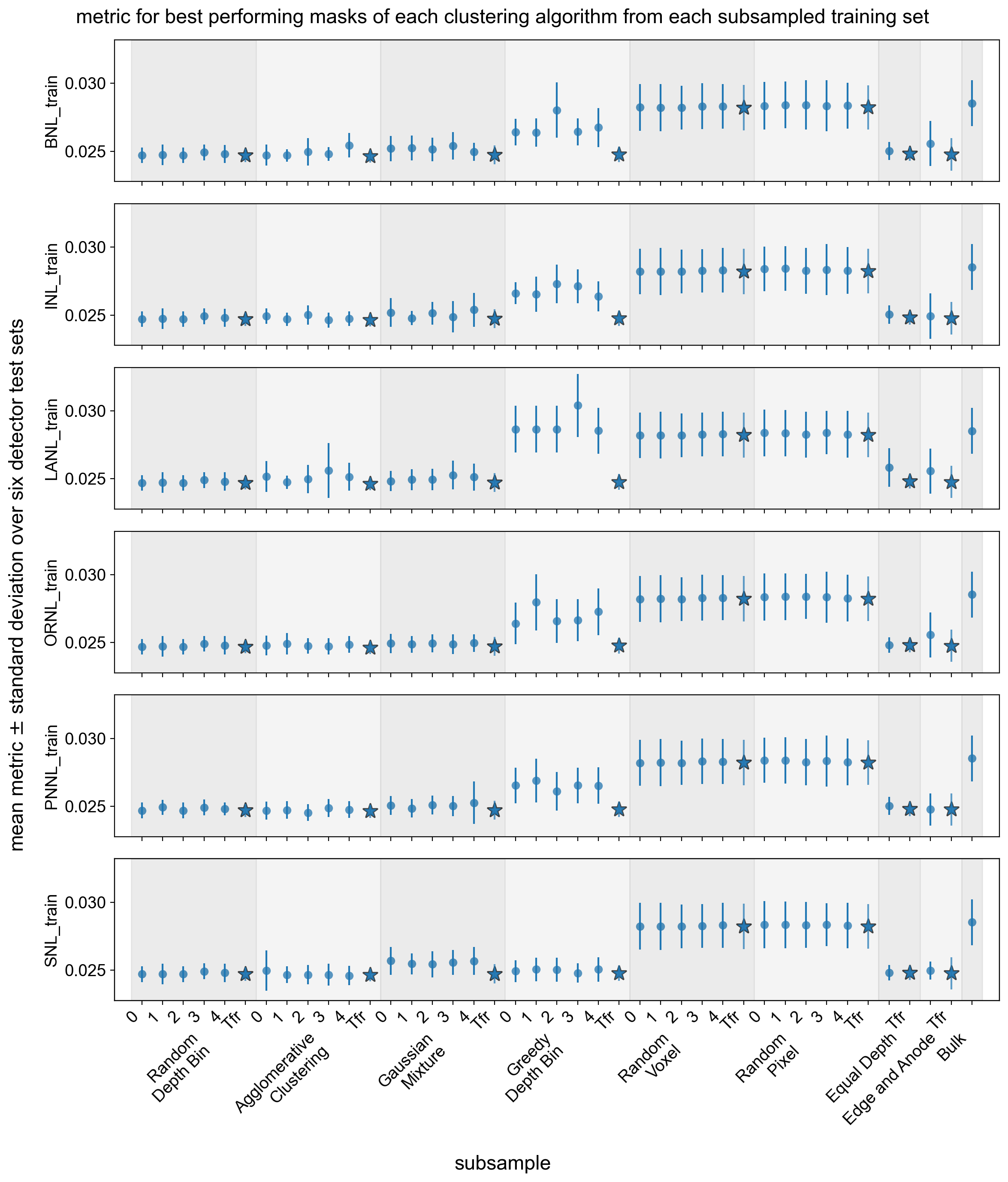}
    \caption{
        Mean and standard deviation of metrics over the six test sets for the best performing mask from each clustering approach from five subsets of each training dataset (circles), with the best transferred (``Tfr'') model performance on the full test set for each clustering algorithm (stars) shown for reference.
        Bars indicate sample standard deviation across the six detectors test sets.
        The edge-and-anode, equal-depth-bin, and bulk masks are constructed independently of the training data, so all training data subsets find the same masks.
    }
    \label{fig:subtrain}
\end{figure*}

Similarly, Fig.~\ref{fig:subtrain} shows the results of the training set subsampling analysis, in which masks were found using five subsamples of each detector training set and performance was averaged over the six test detector datasets.
For brevity, Fig.~\ref{fig:subtrain} shows results only from the best-performing mask within each family of clustering algorithms.
As with the test set subsampling analysis, the subsampled results generally show a narrow distribution of metric values and typically lie close in performance to the full-test-set best-transferred model from that algorithm class, again indicating robustness to statistical variations in the training data and the $5\times$ reduction in training dataset size.
For most algorithms tested, even the subsampled optimized values reliably perform better than the bulk detector.
The exceptions here are the random voxel and random pixel algorithms, which are expected to perform as well as bulk on average, and some of the greedy depth bin algorithms, which appear to struggle under subsampling.

\section{Discussion}\label{sec:discussion}

In general, despite the substantial intra- and inter-detector spatial performance variations observed in Section~\ref{sec:results_characterization}, the transferability studies of Section~\ref{sec:results_transferability} show that high-performing voxel selection masks can be transferred across multiple M400 units to reliably improve performance by excluding on-average poorly-performing detector regions.
In addition, the subsampling analyses of Section~\ref{sec:results_subsampling} suggest that finding such masks is robust to Poisson fluctuations in the training and testing data, as well as a $5\times$ downsample in the datasets used.
Here, we provide some additional interpretations, and discuss some limitations and opportunities for future work.

The Doniach amplitude uncertainty improvement of ${\sim}15\%$ (e.g., from an average of $2.85\%$ to $2.46\%$ in Fig.~\ref{fig:top_metrics}) may, at first glance, seem small, especially given the typical efficiency reduction to ${\sim}30\%$ of bulk.
First, it should be emphasized that efficiency loss is already penalized by the relative uncertainty metric, and that the reduction to ${\sim}30\%$ relative efficiency is more than outweighed by the reduction in the systematic fit error.
Second, as shown in Fig.~\ref{fig:metric_vs_dwell_time}, the optimization ``unlocks'' lower uncertainty values that cannot be achieved by simply dwelling longer with the bulk model, since the systematic fit error component is always present.
Such reductions in uncertainty can directly improve sensitivity to material diversions or isotopic discrepancies in gamma NDA.
In some analyses, such as those in Fig.~5 of Ref.~\cite{vavrek2025bdata}, the optimized model can additionally reach the best-possible bulk model performance in much shorter measurement times; here, the optimized model typically reaches the bulk model performance near to when the bulk performance starts to saturate, offering little potential speedup, but still offering improved performance with longer dwell time.
Reasons for this lack of additional speedup are currently unknown, and could be worth exploring in the future.

The Doniach fit amplitude relative uncertainty used as the performance metric in this work is a useful proxy for more specific, advanced performance metrics that should be constructed for each end-user application.
While this study shows that masks can be transferrable with little performance loss, the specific mask chosen to optimize average performance across detectors in a given application should be recomputed for that application's specific performance target.
In addition, we point out that performance metrics derived from peak fit quality introduce a strong dependence on the exact peak shape used, and can end up rejecting ``good-quality'' detector regions for ostensibly ``small'' model mismatches.
More robust estimates of peak area (net counts) may therefore be preferred in the future.

As has been mentioned in our previous studies~\cite{vavrek2025data, vavrek2025bdata}, further improved mask results may be found by devoting more compute time to a larger parameter sweep.
The parameter sweeps used here (Table~\ref{tab:hyperparam}) were somewhat small and would likely benefit from expansion in a future study with a more application-specific performance metric.
Including data from more than six detectors should also help improve the generalizability of results, reducing the potential impact of overfitting to outlier detectors.

These transferability studies may enable efforts to include well-chosen, transferable binary voxel masks in the M400 data acquisition chain itself, rather than applying the mask in offline postprocessing as is currently done.
Ideally, the mask would produce an optional second readout data stream, such that the raw, unmasked data is never modified and still recorded.
Such efforts should even further improve the efficiency of safeguards NDA, eliminating the need to manually run the voxel selection in postprocessing and then repeat analysis workflows on the postprocessed data.
In this case, the mask(s) should be chosen with the aforementioned recommendations in mind, namely, to use an application-specific performance metric and to conduct larger parameter sweeps, possibly on a larger collection of detectors.

\section{Conclusions}
We have shown that binary voxel selection masks designed to improve the performance of a single M400 gamma spectrometer can be transferred among multiple detectors with little performance loss, despite significant observed intra- and inter-detector performance variability.
These results suggest that a downstream M400 user could acquire measurements from a small number of detectors, determine the on-average best-performing mask from those measurements, and apply that mask to a larger collection of detectors.
This allows for a substantial reduction in the operational complexity of delivering spectroscopic performance optimizations to M400 end-users, ultimately improving safeguards measurement efficiency.

\section*{Acknowledgments}
The U.S.\ Government retains, and the publisher, by accepting the article for publication, acknowledges, that the U.S.\ Government retains a non-exclusive, paid-up, irrevocable, world-wide license to publish or reproduce the published form of this manuscript, or allow others to do so, for U.S.\ Government purposes.

\bibliographystyle{IEEEtran}
\bibliography{biblio}

\end{document}